\documentclass[slac_one]{revtex4}
\usepackage{graphicx}
\usepackage{fancyhdr}
\def\babar{\mbox{\slshape B\kern-0.1em{\small A}\kern-0.1em B\kern-0.1em{\small A\kern-0.2em R}}}

\begin{document}


\noindent BABAR-PROC-08/101\\ SLAC-PUB-13400\\ arXiv:0809.2929[hep-ex]

\title{Hadronic $\boldmath{B}$ Decays to Charm and Charmonium with the BaBar Experiment} 

%

\author{Xavier Prudent}
\affiliation{Laboratoire de Physique des Particules, IN2P3/CNRS et
Universit\'{e} de Savoie, F-74941, Annecy-Le-Vieux, France (for
the BaBar Collaboration)}

\begin{abstract}
The $\babar$ experiment has recorded the decays of more than
465$\times 10^6\ B\bar{B}$ pairs since 1999, and is reaching an
unprecedented precision in the measurement of hadronic $B$ decays.
The following results are presented: tests of QCD factorization
with the decays $B\rightarrow \chi_{c0}K^{*} $, $B\rightarrow
\chi_{c1,2}K^{(*)} $, and $\bar{B}^{0}\rightarrow D^{(*)0}h^{0}$,
$h^0=\pi^0,\ \eta,\ \omega,\ \eta'$, study of the decays to
charmonium $B\rightarrow \eta_c K^{(*)},\ \eta_c(2S) K^{(*)}$ and
$h_cK^{(*)}$, measurement of the mass difference between neutral
and charged $B$'s, measurement of the ``r'' parameters for the
extraction of the CKM angles $\sin(2\beta+\gamma)$ with the decays
$B\rightarrow D_s^{(*)}h,\ h=\pi^-,\ \rho^-,\ K^{(*)+}$, study of
the three-body rare decays $B\rightarrow J/\psi\phi K$, study of
the baryonic decays $\bar{B}^0\rightarrow\Lambda_c^+\bar{p}$,
$B^-\rightarrow\Lambda_c^+\pi^-\bar{p}$, and
$B^-\rightarrow\Lambda_c^+\pi^0\bar{p}$. Except for the results
presented in the sections~\ref{charmo},~\ref{Bmassdiff}
and~\ref{mesratior}, all the given numbers are preliminary.
\end{abstract}

\maketitle

\thispagestyle{fancy}


\section{TESTS OF QCD FACTORIZATION}

Weak decays of hadrons provide a straight access to the parameters
of the CKM matrix and thus to the study of the CP violation. Gluon
scattering in the final state, related with the confinement of
quarks and gluons into hadrons, can modify the decay dynamics and
so must be well understood. In the factorization model~
\cite{ref:BauerStechWirbel,ref:NeubertPetrov}, the
non-factorizable interactions in the final state by soft gluons
are neglected. The matrix element in the effective weak
Hamiltonian of the $B$ decay is then factorized into a product of
independent hadronic currents.

\subsection{Measurement of the branching fractions (BF) of the decays {\boldmath $B\rightarrow \chi_{c0}K^{*}$}~\cite{ref:Kchic0}}

In the factorization model of the decay $b\rightarrow c\bar{c}s$,
the charge conjugation invariance of the current-current operator
forbids the hadronization of $c\bar{c}$ into
$\chi_{c0}$. The branching fractions (BF) of the
decays $B^0\rightarrow \chi_{c0}K^{*0}$ and $B^+\rightarrow
\chi_{c0}K^{*+} $ are measured from exclusive reconstruction using
a data sample of 454$\times 10^6\ B\bar{B}$ pairs in units of $10^{-4}$: $BF(B^0\rightarrow \chi_{c0}K^{*0}) =
1.7 \pm 0.3 \pm 0.2$ and $BF(B^+\rightarrow
\chi_{c0}K^{*+}) = 1.4 \pm 0.5 \pm 0.2$, where
the quoted first errors are statistical and the second are
systematic. The decay $B^0\rightarrow \chi_{c0}K^{*0}$ is observed
with a 8.9 standard deviation (quoted as $\sigma$) significance and an evidence is
found for $B^+\rightarrow \chi_{c0}K^{*+}$ with a 3.6$\sigma$ significance. An upper limit is set for
$BF(B^+\rightarrow \chi_{c0}K^{*+})<2.1$ at 90~$\%$
confidence level (quoted as $CL$). The $B^0\rightarrow \chi_{c0}K^{*0}$ BF does not agree with the zero value expected from
factorization and is about half of the favored mode
$B^0\rightarrow \chi_{c1}K^{*0}$ ($(3.2\pm 0.6)\times
10^{-4}$~\cite{ref:PDG}).

\subsection{Measurement of the BFs of the decays {\boldmath $B\rightarrow\chi_{c1,2}K^{(*)}$}~\cite{ref:Kchic12}}

In the factorization model, no operators exist for the
hadronization of $c\bar{c}$ into $\chi_{c2}$, while the
hadronization to $\chi_{c1}$ is favored. The BFs of the decays $B\rightarrow
\chi_{c1}K^{(*)}$ and $B\rightarrow \chi_{c2}K^{(*)} $ are
measured from exclusive reconstruction using a data sample of 465$\times 10^6\ B\bar{B}$ pairs in units of $10^{-5}$: $BF(B^+\rightarrow\chi_{c1}K^+)=46 \pm
2 \pm 3$, $BF(B^0\rightarrow\chi_{c1}K^0)= 41 \pm
3 \pm 3$, $BF(B^+\rightarrow\chi_{c1}K^{*+})=
27\pm 5 \pm 4$,
$BF(B^0\rightarrow\chi_{c1}K^{*0})= 25\pm 2 \pm 2$, $BF(B^+\rightarrow\chi_{c2}K^+)<1.8~@
90~\%~CL$, $BF(B^0\rightarrow\chi_{c2}K^0)<2.8~@
90~\%~CL$, $BF(B^+\rightarrow\chi_{c2}K^{*+})<12~@
90~\%~CL$, and $BF(B^0\rightarrow\chi_{c2}K^{*0})= 6.4\pm 1.7 \pm
0.5$, where the first quoted errors are statistical
and the second are systematic. The measured values of $BF(B^+\rightarrow\chi_{c1}K^+)$,
$BF(B^0\rightarrow\chi_{c1}K^0)$, and
$BF(B^+\rightarrow\chi_{c1}K^{*+})$ are the most precise to
date. The upper limit on $BF(B^+\rightarrow\chi_{c2}K^+)$ is
improved and evidence for the decay
$B^0\rightarrow\chi_{c2}K^{*0}$ is seen for the first time.

\subsection{Measurement of the BFs of the color-suppressed decays {\boldmath $\bar{B}^{0}\rightarrow D^{(*)0}h^{0}$},
\boldmath{$h^0=\pi^0,\ \eta,\ \omega,\ \eta'$}~\cite{ref:D0h0}}

Previous measurements of the BFs of the color-suppressed decays
$\bar{B}^{0}\rightarrow D^{(*)0}h^{0}$ invalidated the
factorization
model~\cite{ref:Babar2004,ref:Belle2005,ref:Belle2006}.
However more precise measurements are needed to confirm that
result and to constrain the different QCD models:
SCET (Soft Collinear Effective Theory) and pQCD (perturbative
QCD). The BFs are measured from exclusive reconstruction
using a data sample of 454$\times 10^6\ B\bar{B}$ pairs, the measured values are given in
the Table~\ref{tab:table2}.

\begin{table*}[htb]
{\footnotesize \caption{BFs of the decays
$\bar{B}^{0}\rightarrow D^{(*)0}h^{0}$ measured in data.
\label{tab:table2}}
\begin{center}
\begin{tabular}{|l|c|c|}
\hline
$\bar{B}^0$ mode & $(BF\pm\textrm{stat.}\pm\textrm{syst.})\times 10^{-4}$ & Signif. \\
 \hline
$D^0\pi^0$ &  2.78  $\pm$  0.08  $\pm$ 0.20 &  35.5$\sigma$\\
$D^0\eta(\gamma\gamma)$ &  2.34  $\pm$  0.11 $\pm$ 0.17& 26.1$\sigma$ \\
$D^0\eta(\pi\pi\pi^0)$ &  2.51  $\pm$  0.16 $\pm$ 0.17& 20.3$\sigma$\\
  $D^0\eta$ &   2.41  $\pm$  0.09 $\pm$ 0.17  & -  \\
$D^0\omega$ &   2.77  $\pm$  0.13 $\pm$ 0.22&  29.4$\sigma$\\
$D^0\eta'(\pi\pi\eta(\gamma\gamma))$ &  1.29  $\pm$  0.14 $\pm$ 0.09& 14.7$\sigma$\\
$D^0\eta'(\rho^0\gamma)$ &   1.95  $\pm$  0.29 $\pm$ 0.30 & 7.2$\sigma$\\
 $D^0\eta'$ &   1.38  $\pm$  0.12 $\pm$ 0.22  & -  \\
$D^{*0}\pi^0$ &   1.78  $\pm$  0.13 $\pm$ 0.23 & 15.1$\sigma$\\
$D^{*0}\eta(\gamma\gamma)$ &  2.37  $\pm$  0.15 $\pm$ 0.24 & 19.4$\sigma$ \\
$D^{*0}\eta(\pi\pi\pi^0)$ &   2.27  $\pm$  0.23 $\pm$ 0.18& 12.2$\sigma$\\
  $D^{*0}\eta$ &  2.32  $\pm$  0.13 $\pm$ 0.22 & - \\
$D^{*0}\omega$ &  4.44  $\pm$  0.23 $\pm$ 0.61 & 22.3$\sigma$\\
$D^{*0}\eta'(\pi\pi\eta)$ &   1.12  $\pm$  0.26 $\pm$ 0.27 & 8.0$\sigma$\\
$D^{*0}(D^0\pi^0)\eta'(\rho^0\gamma)$ &    1.64  $\pm$  0.53 $\pm$ 0.20 & 3.3$\sigma$\\
 $D^{*0}\eta'$ &   1.29  $\pm$  0.23 $\pm$ 0.23 & -     \\
\hline
\end{tabular}
\end{center}
}
\end{table*}
These results are consistent with the prediction by SCET:
$BF(D^{*0}h^0)/BF(D^0h^0)\sim 1$ for
$h^0\ne\omega$, but marginally consistent with the
predictions by pQCD on the BFs. The measurements
are 3 to 7 times higher than the predictions by the naive
factorization model.

\section{STUDY OF THE {\boldmath$B$}-MESON DECAYS TO {\boldmath $\eta_c K^{(*)},\ \eta_c(2S) K^{(*)},$} AND \boldmath{$h_c\gamma K^{(*)}$}~\cite{ref:etac_hc_etac2S}}\label{charmo}

The $B$ decays to charmonium singlet states $h_c$ and $\eta_c$ are
still poorly known. A better knowledge of the relative abundances
of the various charmonium states allows a deeper understanding of
the underlying strong processes. In the non-relativistic QCD
model, the productions of $\chi_{cJ}$ ($J=0,1,2$) and $h_c$ are
predicted to be comparable in magnitude, however $BF(B\rightarrow
\chi_{c1}K)\sim 3\times 10^{-4}$ and $BF(B^+\rightarrow h_c
K^+)<3.8\times 10^{-5}$. Similarly no exclusive measurements of
the BF of $\eta_c(2S)$ production have been performed. The
knowledge of the mass parameters of the charmonium state $\eta_c$
is pivotal for the models of $c\bar{c}$ spectrum, but the
measurements available so far are in poor agreement with one
another. The large uncertainties on $BF(\eta_c\rightarrow
K\bar{K}\pi)$ and on $BF(\eta_c(2S)\rightarrow K\bar{K}\pi)$ are
cancelled by measuring the ratio by respect to $BF(B^+\rightarrow
\eta_c K^+)$ and $BF(B^+\rightarrow \eta_c(2S) K^+)$. The measured
BF of the $h_c$ and $\eta_c$ productions are measured using
384$\times 10^6\ B\bar{B}$ pairs (with $BF(B^+\rightarrow \eta_c
K^+)=(9.1\pm 1.3)\times 10^{-4}$): $BF(B^0\rightarrow\eta_c
K^{*0}) = ( 5.7 \pm 0.6 (\textrm{stat.}) \pm 0.4 (\textrm{syst.})
\pm 0.8 (\textrm{bf.}) )\times 10^{-4}$, $BF(B^+\rightarrow h_c
K^+)\times BF(h_c\rightarrow\eta_c\gamma) < 4.8\times 10^{-5}~@
90~\%~CL$, and $BF(B^0\rightarrow h_c K^{*0})\times
BF(h_c\rightarrow\eta_c\gamma) <2.2\times 10^{-4}~@ 90~\%~CL$. The
uncertainty noted bf. is related the error on $BF(B^+\rightarrow
\eta_c K^+)$. These are the first upper limits and confirm the
$h_c$ suppression. Using $BF(B^+\rightarrow\eta_c(2S)K^+)=(3.4\pm
1.8)\times 10^{-4}$, the upper limit on BF for $\eta_c(2S)$
production is $BF(B^0\rightarrow \eta_c(2S)K^{*0}) < 3.9 \times
10^{-4}~@90~\%\ CL$. Using $BF(B^+\rightarrow\eta_c K^+)\times
BF(\eta_c\rightarrow K\bar{K}\pi)=(6.88\pm
0.77^{+0.55}_{-0.66})\times 10^{-4}$, the first measurement is
reported for $BF(\eta_c(2S)\rightarrow K\bar{K}\pi) = (1.9 \pm
0.4\textrm{(stat.)} \pm 0.5\textrm{(syst.)} \pm 1.0\textrm{(bf.)})
$. Both the mean and width of the $\eta_c$ mass distribution are
extracted: $m(\eta_c) = (2985.8 \pm 1.5\textrm{(stat.)} \pm
3.1\textrm{(syst.)})~\textrm{MeV}/c^2$, $\Gamma(\eta_c) =
(36.3^{+3.7}_{-3.6}\textrm{(stat.)} \pm
4.4\textrm{(syst.)})~\textrm{MeV}$, which are in agreement with
the previous $\babar$ measurements.

\section{MEASUREMENT OF THE MASS DIFFERENCE {\boldmath $m(B^0)-m(B^+)$}~\cite{ref:BmassDiff}}\label{Bmassdiff}

The measurement of the mass difference $\Delta m_B =
m(B^0)-m(B^+)$ probes the Coulomb contributions to the quark
structure, which affect the relative production rates of
$\Upsilon(4S)\rightarrow B^0\bar{B}^0$ and
$\Upsilon(4S)\rightarrow B^+B^-$.
The decay modes $B^0\rightarrow J/\psi K^+\pi^-$ and
$B^+\rightarrow J/\psi K^+$ with $J/\psi\rightarrow e^+e^-,\
\mu^+\mu^-$, are reconstructed exclusively using 230 million$\times 10^6\ B\bar{B}$ pairs. The mass difference $\Delta m_B$ is then
computed as:

\begin{equation}
\Delta m_B = - \Delta p^* \times
\frac{p^*(B^0)+p^*(B^+)}{(m(B^0)+m(B^+))\cdot c^2},
\end{equation}
where $p^*$ is the momentum in the $\Upsilon(4S)$ rest frame. The
measured value is $\Delta m_B = (0.33 \pm 0.05\textrm{(stat.)} \pm
0.03\textrm{(syst.)} )~\textrm{MeV}/c^2$, which excludes the null
value at the 5$\sigma$ level.

\section{MEASUREMENT OF THE BFs OF {\boldmath $B^0\rightarrow D_s^{(*)+}\pi^-$}, {\boldmath $B^0\rightarrow D_s^{(*)+}\rho^-$}, AND {\boldmath $B^0\rightarrow
D_s^{(*)-}K^+$}~\cite{ref:sin2betaG}}\label{mesratior}

The quantity $\sin(2\beta+\gamma)$, with the CKM parameters
$\beta$ and $\gamma$, can be measured from the study of the time
evolution of the doubly-Cabibbo and CKM-suppressed decays
$B^0\rightarrow D^{(*)-}\pi^+$ and $B^0\rightarrow
D^{(*)-}\rho^+$. That study requires the knowledge of the ratios
of the decay amplitudes $r(D^{(*)}\pi)= |A(B^0\rightarrow
D^{(*)+}\pi^-)/A(B^0\rightarrow D^{(*)-}\pi^+)|$,
which cannot be directly measured. Assuming $SU(3)$ flavor
symmetry, $r(D^{(*)}\pi)$ can be related to the decay
$B^0\rightarrow D_s^{(*)+}\pi^-$:
\begin{equation}\label{eq:SU3}
r(D^{(*)}\pi)=\tan(\theta_c)\frac{f_{D^{(*)}}}{f_{D_s^{(*)}}}\sqrt{\frac{BF(B^0\rightarrow
D_s^{(*)+}\pi^-)}{BF(B^0\rightarrow D^{(*)-}\pi^+)}},
\end{equation}
where $\theta_c$ is the Cabibbo angle, and
$f_{D^{(*)}}/f_{D_s^{(*)}}$ is the ratio of $D^{(*)}$ and
$D_s^{(*)}$ meson decay constants.

The contribution from W-exchange diagrams are evaluated from the
study of $B^0\rightarrow D_s^{(*)-}K^+$, which proceeds through a
W-exchange diagram only.

Using 381$\times 10^6\ B\bar{B}$ pairs, the measured BFs are (in units of
$10^{-5}$): $BF(D_s^+\pi^-) = 2.5 \pm 0.4 \pm
0.2$, $BF(D_s^{*+}\pi^-) = 2.6^{+0.5}_{-0.4} \pm 0.2$,
$BF(D_s^+\rho^-) < 2.4~@ 90~\%~CL$, $BF(D_s^{*+}\rho^-) =
4.1^{+1.3}_{-1.2} \pm 0.4$, $BF(D_s^-K^-) = 2.9 \pm 0.4 \pm 0.2$,
$BF(D_s^{*-}K^+) = 2.4 \pm 0.4 \pm 0.2$, $BF(D_s^-K^{*+}) =
3.5^{+1.0}_{-0.9} \pm 0.4$, and $BF(D_s^{*-}K^{*+}) =
3.2^{+1.4}_{-1.2} \pm 0.4$.

The measured longitudinal fractions are:
$f_L(D_s^{*+}\rho^-) = 0.84^{+0.26}_{-0.28} \pm 0.13$ and
$f_L(D_s^{*-}K^{*+}) = 0.92^{+0.37}_{-0.31} \pm 0.07$. The values
of $r(D^{(*)}\pi)$ are computed with Equation~(\ref{eq:SU3}): $r(D\pi) =
(1.78^{+0.14}_{-0.13}  \pm 0.08 \pm 0.10 (\textrm{th.}))~\%$,
$r(D^{*}\pi) = (1.81^{+0.16}_{-0.15}  \pm 0.09  \pm 0.10
(\textrm{th.}))~\%$, $r(D\rho) = (0.71^{+0.29}_{-0.27}
 \pm 0.10  \pm 0.04
(\textrm{th.}))~\%$, and $r(D^{*}\rho) = (1.45^{+0.23}_{-0.22}
 \pm 0.12  \pm 0.08
(\textrm{th.}))~\%$. The quoted first errors are statistical and
the second are systematic. The errors denoted th. are related to
the theoretical uncertainties.

\section{MEASUREMENT OF THE BFs OF THE RARE DECAYS {\boldmath $B\rightarrow J/\psi\phi K$}~\cite{ref:JPsiPhiK}}

Many charmonium-like resonances were discovered recently and more
are expected in the $J/\psi\phi$ decay
channel. The decays $B^0\rightarrow J/\psi\phi
K^0$ and $B^+\rightarrow J/\psi\phi K^+$ were exclusively
reconstructed using 433$\times 10^6\ B\bar{B}$ pairs. The measured BFs
are: $BF(B^0\rightarrow J/\psi\phi K^0) = (
5.40 \pm 1.20(\textrm{stat.}) \pm 0.40(\textrm{syst.}) )\times
10^{-5}$ and $BF(B^+\rightarrow J/\psi\phi K^+) = (5.81 \pm
0.73(\textrm{stat.}) \pm 0.29(\textrm{syst.}) )\times 10^{-5}$.
The study of the $J/\psi\phi$ mass spectrum is on-going.

\section{STUDY OF BARYONIC \boldmath{$B$} DECAYS}

Baryonic decays of $B$ mesons provide a laboratory for searches for
excited charm baryon states and for the investigation of the
dynamics of 3-body decays.

\subsection{Study of the decays
{\boldmath $\bar{B}^0\rightarrow\Lambda_c^+\bar{p}$} and
\boldmath{$B^-\rightarrow\Lambda_c^+\pi^-\bar{p}$}~\cite{ref:Lambdacppi}}

The BFs of the decay channels
$\bar{B}^0\rightarrow\Lambda_c^+\bar{p}$ and
$B^-\rightarrow\Lambda_c^+\pi^-\bar{p}$ are measured from
exclusive reconstruction using 383$\times 10^6\ B\bar{B}$ pairs:
$BF(\bar{B}^0\rightarrow\Lambda_c^+\bar{p}) = (1.89 \pm
0.21(\textrm{stat.}) \pm 0.06(\textrm{syst.}) \pm
0.49(\textrm{bf.}))\times 10^{-5}$ and $BF(B^-\rightarrow\Lambda_c^+\pi^-\bar{p})
= (3.38 \pm 0.12(\textrm{stat.}) \pm 0.12(\textrm{syst.}) \pm
0.88(\textrm{bf.}))\times 10^{-4} $, where the error denoted bf. is related to the
uncertainty on $BF(\Lambda_c^+\rightarrow p K^-\pi^+)$. One notices
an enhancement of the 3-body channel by a factor 15 by respect to
the 2-body channel. An enhancement is seen in the Dalitz plot of
$B^-\rightarrow\Lambda_c^+\pi^-\bar{p}$ at the threshold of the
phase space in $m^2(\Lambda_c\bar{p})$. Such threshold enhancement
has been seen in other baryon-antibaryon decay modes and is thus
expected to be a dynamical effect rather than a
resonance. Three resonances are investigated
in the $\Lambda_c\pi$ mass spectrum: $\Sigma_c(2455)^0$,
$\Sigma_c(2520)^0$ and $\Sigma_c(2800)^0$. The relative BFs
measured are:
$BF(B^-\rightarrow\Sigma_c(2455)^0\bar{p})/BF(\bar{B}^0\rightarrow\Lambda_c^+\pi^-\bar{p})
= (12.3 \pm 1.2(\textrm{stat.}) \pm 0.8(\textrm{syst.}))\times
10^{-2}$,
$BF(B^-\rightarrow\Sigma_c(2800)^0\bar{p})/BF(\bar{B}^0\rightarrow\Lambda_c^+\pi^-\bar{p})
= (11.7 \pm 2.3(\textrm{stat.}) \pm 2.4(\textrm{syst.}))\times
10^{-2}$, and
$BF(B^-\rightarrow\Sigma_c(2520)^0\bar{p})/BF(\bar{B}^0\rightarrow\Lambda_c^+\pi^-\bar{p})
< 0.9\times 10^{-2}~@ 90~\%~CL$. No signal is seen for
$\Sigma_c(2520)^0$.

The parameters of the mass distributions of these resonances are
extracted: $m(\Sigma_c(2455)^0) = 2454.0 \pm
0.2~\textrm{MeV}/c^2$, $\Gamma(\Sigma_c(2455)^0) = 2.6\pm
0.5~\textrm{MeV}$, $m(\Sigma_c(2800)^0) = 2846.0 \pm
8.0~\textrm{MeV}/c^2$, and $\Gamma(\Sigma_c(2800)^0) =
86^{+33}_{-22}~\textrm{MeV}$. The measured mass for
$\Sigma_c(2800)^0$ is 3$\sigma$ higher than the resonance seen by
Belle~\cite{ref:BelleSigma2800}, which may indicate a new $J=1/2$
state. The angular distribution of
$B^-\rightarrow\Sigma_c(2455)^0\bar{p}$ is consistent with a spin
of $J=1/2$ for $\Sigma_c(2455)^0$ and the hypothesis $J=3/2$ is
rejected at the $> 4\sigma$ level.

\subsection{Study of the decay {\boldmath$\bar{B}^0\rightarrow\Lambda_c^+\pi^0\bar{p}$}}

This channel is the isospin counterpart of
$B^-\rightarrow\Lambda_c^+\pi^-\bar{p}$ and has never been observed.
The BF is measured in the restrictive phase space
$m(\Lambda_c^+\pi^0)>3.0~\textrm{GeV}/c^2$ and so does not include
contributions from $\Sigma_c(2455,2520,2800)^0$ resonances. An
enhancement, similar to the one seen for
$B^-\rightarrow\Lambda_c^+\pi^-\bar{p}$, is seen at the threshold
of the phase space of the $\Lambda_c^+\bar{p}$ mass spectrum.
Using 467$\times 10^6\ B\bar{B}$ pairs the measured BF is:
$BF(\bar{B}^0\rightarrow\Lambda_c^+\pi^0\bar{p}) = (1.61 \pm
0.26(\textrm{stat.}) \pm 0.13(\textrm{syst.})  \pm
0.42(\textrm{bf.}))\times 10^{-4}$, where the error denoted bf. is
related to the uncertainty on $BF(\Lambda_c^+\rightarrow p
K^-\pi^+)$.

\end{document}